\documentstyle[12pt]{article}

\catcode`\@=11
\def\marginnote#1{}
\newcount\hour
\newcount\minute
\newtoks\amorpm
\hour=\time\divide\hour by60
\minute=\time{\multiply\hour by60 \global\advance\minute
by-\hour}\edef\standardtime{{\ifnum\hour<12
\global\amorpm={am}%
        \else\global\amorpm={pm}\advance\hour by-12 \fi
        \ifnum\hour=0 \hour=12 \fi
        \number\hour:\ifnum\minute<10
0\fi\number\minute\the\amorpm}}
\edef\militarytime{\number\hour:\ifnum\minute<10
0\fi\number\minute}

\def\draftlabel#1{{\@bsphack\if@filesw {\let\thepage\relax
   \xdef\@gtempa{\write\@auxout{\string
      \newlabel{#1}{{\@currentlabel}{\thepage}}}}}\@gtempa
   \if@nobreak \ifvmode\nobreak\fi\fi\fi\@esphack}
        \gdef\@eqnlabel{#1}}
\def\@eqnlabel{}
\def\@vacuum{}
\def\draftmarginnote#1{\marginpar{\raggedright\scriptsize\tt#1}}
\def\draft{\oddsidemargin -.5truein
        \def\@oddfoot{\sl preliminary draft \hfil
        \rm\thepage\hfil\sl\today\quad\militarytime}
        \let\@evenfoot\@oddfoot \overfullrule 3pt
        \let\label=\draftlabel
        \let\marginnote=\draftmarginnote

\def\@eqnnum{(\theequation)\rlap{\kern\marginparsep\tt\@eqnlabel}%
\global\let\@eqnlabel\@vacuum}  }

\def\numberbysection{\@addtoreset{equation}{section}
        \def\theequation{\thesection.\arabic{equation}}}

\def\underline#1{\relax\ifmmode\@@underline#1\else
 $\@@underline{\hbox{#1}}$\relax\fi}

\catcode`@=12
\relax

\numberbysection

\advance \topmargin by -\headheight
\advance \topmargin by -\headsep

\evensidemargin \oddsidemargin
\def\br{\begin{eqnarray}}
\def\er{\end{eqnarray}}
\def\be{\begin{equation}}
\def\ee{\end{equation}}

\def\({\left(}
\def\){\right)}

\newcommand{\bi}[1]{\bibitem{#1}}

\relax

\def\b{\beta}

\def\d{\delta}
\def\D{\Delta}

\def\g{\gamma}

\def\pa{\partial}

\def\s{\sigma}

\def\tp0{\Theta_{+}^{(0)}}
\def\tm0{\Theta_{-}^{(0)}}

\def\f#1#2#3 {f^{#1#2}_{#3}}

\def\win1{{\sf w_{1+\infty}}}

\def\Win1{{\sf W_{1+\infty}}}

\def\rlx{\relax\leavevmode}
\def\inbar{\vrule height1.5ex width.4pt depth0pt}
\def\IZ{\rlx\hbox{\sf Z\kern-.4em Z}}
\def\IR{\rlx\hbox{\rm I\kern-.18em R}}
\def\IC{\rlx\hbox{\,$\inbar\kern-.3em{\rm C}$}}
\def\IN{\rlx\hbox{\rm I\kern-.18em N}}
\def\IO{\rlx\hbox{\,$\inbar\kern-.3em{\rm O}$}}
\def\IP{\rlx\hbox{\rm I\kern-.18em P}}
\def\IQ{\rlx\hbox{\,$\inbar\kern-.3em{\rm Q}$}}
\def\IF{\rlx\hbox{\rm I\kern-.18em F}}
\def\IG{\rlx\hbox{\,$\inbar\kern-.3em{\rm G}$}}
\def\IH{\rlx\hbox{\rm I\kern-.18em H}}
\def\II{\rlx\hbox{\rm I\kern-.18em I}}
\def\IK{\rlx\hbox{\rm I\kern-.18em K}}
\def\IL{\rlx\hbox{\rm I\kern-.18em L}}
\def\one{\hbox{{1}\kern-.25em\hbox{l}}}
\def\0#1{\relax\ifmmode\mathaccent"7017{#1}%
B        \else\accent23#1\relax\fi}

\def\EPC#1#2#3{{\sl Eur. Phys. J.} {\bf C#1} (#2) #3}
                \def\JHEP#1#2#3{{\sl JHEP} {\bf#1} (#2) #3}
                
                \def\NPB#1#2#3{{\sl Nucl. Phys.} {\bf B#1} (#2) #3}
                
                \def\CMP#1#2#3{{\sl Commun. Math. Phys.} {\bf #1} (#2) #3}
                \def\PRD#1#2#3{{\sl Phys. Rev.} {\bf D#1} (#2) #3}

                \def\JMP#1#2#3{{\sl J. Math. Phys.} {\bf #1} (#2) #3}

                \def\AoP#1#2#3{{\sl Annals Phys.} {\bf #1} (#2) #3}

                \def\PR#1#2#3{{\sl Phys. Reports} {\bf #1} (#2) #3}

                \def\JPA#1#2#3{{\sl J. Physics} {\bf A#1} (#2) #3}

                \def\b{\beta}

                \def\d{\delta}
                \def\D{\Delta}
                
                \def\g{\gamma}

                \def\/{\frac}

                \def\pa{\partial}

                \def\s{\sigma}

                \def\({\Big(}
                \def\){\Big)}
                \def\[{\Big[}
                \def\]{\Big]}

                \def\rlx{\relax\leavevmode}
                \def\inbar{\vrule height1.5ex width.4pt depth0pt}
                \def\IZ{\rlx\hbox{\sf Z\kern-.4em Z}}
                \def\IR{\rlx\hbox{\rm I\kern-.18em R}}
                \def\IC{\rlx\hbox{\,$\inbar\kern-.3em{\rm C}$}}
                \def\IN{\rlx\hbox{\rm I\kern-.18em N}}
                \def\IO{\rlx\hbox{\,$\inbar\kern-.3em{\rm O}$}}
                \def\IP{\rlx\hbox{\rm I\kern-.18em P}}
                \def\IQ{\rlx\hbox{\,$\inbar\kern-.3em{\rm Q}$}}
                \def\IF{\rlx\hbox{\rm I\kern-.18em F}}
                \def\IG{\rlx\hbox{\,$\inbar\kern-.3em{\rm G}$}}
                \def\IH{\rlx\hbox{\rm I\kern-.18em H}}
                \def\II{\rlx\hbox{\rm I\kern-.18em I}}
                \def\IK{\rlx\hbox{\rm I\kern-.18em K}}
                \def\IL{\rlx\hbox{\rm I\kern-.18em L}}
                \def\one{\hbox{{1}\kern-.25em\hbox{l}}}
                \def\0#1{\relax\ifmmode\mathaccent"7017{#1}%
                B        \else\accent23#1\relax\fi}
                
\begin{document}
                \begin{center}
                  {\large\bf Bosonization and the generalized Mandelstam operators}
                \end{center}

                \begin{center}
                Harold Blas
                \par \vskip .1in
ICET-Universidade Federal de Mato Grosso (UFMT)\\
 Av. Fernando Correa, s/n, Coxip\'o \\
78060-900, Cuiab\'a - MT - Brazil
\end{center}

\begin{abstract}
\vspace{.2 cm}

The generalized massive Thirring model (GMT) with $N_{f}(=$number
of positive roots of $su(n)$) fermion species is bosonized in the
context of the functional integral and operator formulations and
shown to be equivalent to a generalized sine-Gordon model (GSG)
with $N_{f}$ interacting soliton species. The generalized
Mandelstam-Halpern soliton operators are constructed and the
fermion-boson mapping is established through a set of generalized
bosonization rules in a quotient positive definite Hilbert space
of states. Each fermion species is mapped to its corresponding
soliton in the spirit of particle/soliton duality of Abelian
bosonization. The examples of $su(3)$ and $su(4)$ are presented.
\end{abstract}





                \section{Introduction}
The transformation of Fermi fields into Bose fields, called  {\sl
bosonization}, provided in
                the last years a powerful tool to obtain non-perturbative information
                in two-dimensional field theories \cite{abdalla}. The
                Abelian and non-Abelian bosonizations have been derived in \cite{coleman} and \cite{witten1}, respectively. In
                 the non-abelian
                 developments the appearance of solitons in the
                 bosonized model, which generalizes the
                 sine-Gordon solitons, to our knowledge has not been fully
                 explored. The interacting
                 multi-flavor massive fermions deserves a
                 consideration in the spirit of the particle/soliton duality of the
                 Abelian bosonization.

               In this context, an  important question is related to the
                multi-flavor extension of the well known massive Thirring (MT)
                and sine-Gordon relationship (SG)\cite{coleman}. In \cite{jmp, jhep} it has been shown
                 that the   generalized  massive Thirring model (GMT) is equivalent to the
                generalized sine-Gordon model (GSG) at the classical level; in particular, the mappings between spinor bilinears of the GMT theory and exponentials of the GSG fields were
                established on shell and the various soliton/particle correspondences were
                uncovered.

In \cite{jhepc} the bosonization of the GMT model has been
performed following a hybrid of the operator and  functional
formalisms. This approach introduces a redundant Bose field
algebra containing some un-physical degrees of freedom
\cite{belvedere}. The redundant Bose fields constitute a set of
pairwise  massless fields quantized
                with opposite metrics. In the GMT cases, under
                consideration here, these features are reproduced according to an
                affine $su(n)$ Lie algebraic construction.

                A positive definite Hilbert space of states ${\cal H}$ is identified as
                a quotient space in the Hilbert space hierarchy emerging in the
                bosonization process. One
                has
                that each GMT fermion is bosonized in terms of a Mandelstam
                ``soliton'' operator and a spurious exponential field with zero scale
                dimension, this   spurious field behaves as an identity in the Hilbert space
                 ${\cal H}$
                and, so, has no physical effects. Afterwards, a set of generalized
                bosonization rules are established mapping the GMT fermion bilinears into
                the corresponding operators composed of the GSG boson fields.

                \section{Functional integral and operator approaches}

The two-dimensional massive Thirrring model with current-current
                interactions of $N_{f}$ (Dirac) fermion species is defined by the
                Lagrangian
                density
                \begin{eqnarray}
                \label{thirring1}
                \frac{1}{k'}{\cal L}_{GMT}[\psi^{j},\overline{\psi}^{j}]=
                \sum_{j=1}^{N_{f}} \{i\bar{\psi}^{j}\gamma^{\mu}\pa_{\mu}\psi^{j}
                        - m^{j}\,\,{\overline{\psi}}^{j}\psi^{j}\}\,
                        \,- \frac{1}{4}
                \sum_{k,\,l=1}^{N_{f}}\[\hat{G}_{kl}\,J_{k}^{\mu} J_{l\, \mu }\],
                \end{eqnarray}
                where the $m^{j}$'s are the mass parameters, the overall coupling $k'$
                has been introduced for later purposes, the currents are defined by
                $J_{j}^{\mu}\,=\,\bar{\psi}^{j}\gamma^{\mu}\psi^{j}$, and the coupling
                constant parameters are represented by a non-degenerate
                $N_{f}\,$x$\,N_{f}$
                symmetric matrix $
                \hat{G}=\hat{g} {\cal G} \hat{g},\,\,\,\,\hat{g}_{ij}\,=\,g_{i}
                \d_{ij},\,\,\,\,{\cal G}_{jk}={\cal G}_{kj}.$

                The GMT model (\ref{thirring1}) is related to the weak coupling sector
                of the $su(n)$ ATM theory in the classical treatment of Refs.
                \cite{jmp, jhep}. We shall consider the special cases of
                $su(n)$ ($n=3, 4$). In the $n=3$ case the currents at the quantum
                level must satisfy
                \br
                \label{currelat}
                J_{3}^{\mu}\,=\, \hat{\d}_{1} J_{1}^{\mu} + \hat{\d}_{2} J_{2}^{\mu},
                \er
                where the $\hat{\d}_{1,\,2}$ are some parameters related to
                the   couplings $\hat{G}_{kl}$. Similarly, in the $n=4$ case the currents at
                the quantum level satisfy
                \br
                \label{currelatsu4}
                J_{4}^{\mu}\,=\, \hat{\s}_{41} J_{1}^{\mu} + \hat{\s}_{42}
                J_{2}^{\mu},\,\,\,\,
J_{5}^{\mu}\,=\, \hat{\s}_{51} J_{1}^{\mu} + \hat{\s}_{53}
J_{3}^{\mu},\,\,\,\, J_{6}^{\mu}\,=\, \hat{\s}_{62} J_{2}^{\mu} +
\hat{\s}_{63} J_{3}^{\mu},
                \er
where the $\hat{\s}_{ia}$'s (i=4,5,6; a=1,2,3) are related to the
couplings $\hat{G}_{kl}$.

                The cases under consideration $n=3,4$ correspond to $N_{f}=3, 6$, respectively; however,
                most of the construction below is valid for $N_{f}>6$. In the hybrid approach the Thirring fields are written in terms of the
                ``generalized''
                 Mandelstam ``soliton'' $\Psi^{j}(x)$ and $\s_{j}$
                 fields
                \br
                \label{map2}
                \psi^{j}(x) = \Psi^{j}(x) \s^{j},\,\,\,\,\, j=1,2,3,...,N_{f};
                \er
                where
                \br
                \label{fermion1}
                \Psi^{j}(x)& =& (\frac{\mu}{2\pi})^{1/2}\, K_{j} \,\, e^{-i\pi \g_{5}/4}
                :e^{-i\( \frac{\b_{j}}{2} \g_{5} \Phi^{j}(x) +
                \frac{2\pi}{\b_{j}}\int_{x^{1}}^{+\infty}\dot{\Phi^{j}}(x^{0},z^{1})
                dz^{1}
                \)}:\\
                \label{fermion2}
                \s^{j}&=&
                e^{\frac{i}{2}\(\eta_{j}-\frac{\sqrt{4\pi}}{\D_{j}}\widetilde{\xi}_{j}\)}\\
                \label{fermion3}
                &=& e^{-\frac{i}{2} g_{j} \ell^{j}}.
                \er

                In (\ref{fermion1}) the factor $K_{j}$ makes the fields anti-commute for different
                flavors \cite{frishman}.

The Lagrangian in terms of
                purely bosonic fields
                becomes
                \begin{eqnarray}
                \frac{1}{k'}{\cal L}_{eff}^{'}&=& \sum_{j, k = 1}^{N_{f}} \frac{1}{2}
                \[C_{jk}\, \pa_{\mu} \Phi_{j}\pa^{\mu} \Phi_{k} + E_{jk}\,  \pa_{\mu} \xi_{j}\pa^{\mu}
                \xi_{k}
                \label{lageff1}
                + F_{jk}\, \pa_{\mu} \eta_{j}\pa^{\mu} \eta_{k} \]  + \sum_{j=1}^{3}  M^{j} \mbox{cos}(\Phi_{j}),
                \end{eqnarray}
                where\, $C_{jk},  D_{jk}, E_{jk}$ and $M^{j}$ \,
                are some parameters such that the fields $\xi_{j}$ and $\eta_{j}$ are quantized with opposite metric.

                Notice that each $\Psi^{j}$ is written
in terms of a non-local
                expression of  the corresponding bosonic field $\Phi^{j}$ and
                the appearance of the couplings $\b_{j}$ in (\ref{fermion1})
                in  the same form as in the standard sine-Gordon construction of the
                Thirring fermions \cite{coleman}; so,  one can refer the fermions
                $\Psi^{j}(x)$ as  generalized SG Mandelstam-Halpern soliton operators.

\subsection{The $su(3)$ case}

The generalized massive   Thirring
                model (GTM) (\ref{thirring1}) with three fermion species, satisfying   the
                currents constraint (\ref{currelat}), bosonizes to the generalized
                sine-Gordon model (GSG) (\ref{lageff1}) with three boson fields
                satisfying  the linear constraint $\b_{3}\Phi_{3}=\b_{1}\Phi_{1}\d_{1}+\d_{2}\b_{2}\Phi_{2}$,
                by means of the ``generalized'' bosonization rules
                \br
                \label{bosoni1}
                i\bar{\psi}^{j}\g^{\mu}\pa_{\mu}\psi^{j} &=& \frac{1}{2}(1-\rho_{j})
                (\pa_{\mu}\Phi^{j})^2,\,\,\,\,j=1,2,3;\\
                \label{bosoni2}
                 m_{j} \bar{\psi}^{j}\psi^{j}&=& M_{j}\, \mbox{cos}\(\b_{j}
                \Phi^{j}\),\,\,\,\,\,\,\,\,\b_{j}^2=\frac{4\pi}{1+\frac{g_{j}^2}{\pi}\frac{1}{4
                {\cal G}_{lm}^{j}}} \\
                \label{bosoni3}
                \bar{\psi}^{j}\g^{\mu} \psi^{j}&=& -
                \frac{\b_{j}}{2\pi}\,\epsilon^{\mu\nu}\pa_{\nu}\Phi_{j},
                \er
                where $\rho_{j}$ can be written in terms of $g_{j}$ and the correlation functions on the right hand sides must be
                understood to be computed in a positive definite quotient Hilbert space
                of
                states ${\cal H} \sim \frac{{\cal H}'}{{\cal H}_{o}}$ \cite{jhepc}.

\subsection{The $su(4)$ case}

In the $su(4)$ case \cite{nova2} one has the constraints
$\Phi_{4}= \s_{41} \Phi_{1} + \s_{42} \,
                \Phi_{2},\,\,\,\,\Phi_{5}= \s_{51} \Phi_{1} + \s_{53} \,
                \Phi_{3},\,\,\,\,\Phi_{6}= \s_{62} \Phi_{2} + \s_{63} \,
                \Phi_{3}$, \, with $\s_{ij}$ being some parameters. The bosonization
                rules become
                  \br
                \label{bosoni1su4}
                \label{currents11su4}
                \bar{\psi}^{j}\g^{\mu}\psi^{j} &=& -
                \frac{\hat{\b}_{i}\,\epsilon^{\mu\nu}\pa_{\nu}\Phi_{i}}{\pi {\cal M}_{i}^{-} {\cal
                M}_{i}^{+}- (\frac{z_{i}}{2})\, (g_{i})^2\,{\cal M}_{i}^{+}}
                \,\,\,\,i=1,2,3,...,6;\\
                i\bar{\psi}^{j}\g^{\mu}\pa_{\mu}\psi^{j} &=& \frac{1}{2}(1-\hat{\rho}_{j})
                (\pa_{\mu}\Phi^{j})^2,\,\,\,\,j=1,2,3...,6;\\
                \label{bosoni2su4}
                 m_{j} \bar{\psi}^{j}\psi^{j}&=& M_{j}\, \mbox{cos}\(\hat{\b}_{j}
                \Phi^{j}\). \er

                The parameters ${\cal M}_{i}^{\pm}$, $\hat{\b}_{j}$ and
                $\hat{\rho}_{j}$ can be expressed in terms of $g_{i}$, ${\cal
                G}_{ij}$. The $z_{i}$'s  are regularization
                parameters.

                \section{Discussions}

                Using the mixture of the functional integral and operator formalisms we
                have considered the bosonization of the multiflavour
                 GMT model with $N_{f}$ [=number of positive roots of $su(n)$] species.
                 The sets of free bosonic fields ($\xi_{j},\,
                \eta_{j}$) are quantized with opposite metrics and their contributions
                are
                essential in order to define the correct Hilbert space of states and
                the relevant fermion-boson mappings. One must emphasize that the
                classical properties of the ATM model \cite{jmp, jhep, nucl} motivated the
                various insights considered in the bosonization procedure of the GMT model
performed in this
                work. The form of the quantum GSG model (\ref{lageff1}) is similar to its
                classical counterparts in \cite{jmp, jhep}, except for the field
                renormalizations and the relevant quantum corrections to the coupling
                constants. The bosonization results presented in this talk can be useful to
                study particle/soliton
                duality and confinement in multi-fermion two-dimensional
                models, extending the results of \cite{nucl1, tension}.

                {\sl Acknowledgments}

               The author thanks the organizers of the V-SILAFAE and the hospitality of IMPA,
               CBPF and ICET-UFMT. This work was supported by FAPEMAT-CNPq.


\begin{thebibliography}{**}
                \bibitem{abdalla}
                E. Abdalla, M.C.B. Abdalla and K.D. Rothe, Non-perturvative methods in
                two-dimensional quantum field theory, 2nd edition (World Scientific,
                Singapore, 2001).
\bibitem{coleman}
                S. Coleman, \PRD{11}{1975}{2088};\\
                S. Mandelstam, \PRD{11}{1975}{3026}.\bibitem{witten1}
E. Witten, \CMP{92}{1984}{455}.
                \bibitem{jmp}
                J. Acosta, H. Blas, \JMP{43}{2002}{1916};\\
H. Blas, to appear in {\sl Progress in Soliton Research} (Nova
Science Publishers, 2004), hep-th/0407020.
                \bi{jhep}
                H. Blas, \JHEP{0311}{2003}{054}.
\bibitem{jhepc}
                H. Blas, \EPC{37}{2004}{251}.

\bibitem{belvedere}
                L. V. Belvedere and R. L. P. G. Amaral,
                \PRD{62}{2000}{065009};\\
                L.V. Belvedere, \JPA{33}{2000}{2755}.\bibitem{frishman}
Y. Frishman and J. Sonnenschein,
\PR{223}{1993}{309}.\bibitem{nova2} H. Blas, invited paper for
{\sl Progress in Boson Research} (Nova Science Publishers, 2005),
hep-th/0409269.
                \bibitem{nucl}
                H. Blas, \NPB{596}{2001}{471};\\
                H. Blas and B.M. Pimentel, \AoP{282}{2000}{67}; see also  hep-th/0005037.
                \bibitem{nucl1}
                H. Blas and L.A. Ferreira, \NPB{571}{2000}{607}.
                \bibitem{tension}
                H. Blas, \PRD{66}{2002}{127701}; see also hep-th/0005130.
\end{thebibliography}
\end{document}